\documentclass[aps,prc,preprint,amsmath,amssymb,showpacs,preprintnumbers,superscriptaddress]{revtex4-1}
\usepackage{CJK}
\usepackage{graphicx}
\usepackage{dcolumn}
\usepackage{bm}
\usepackage{mathrsfs}   
\usepackage{color,xcolor}  
\usepackage{multirow}  
\usepackage{booktabs} 

\usepackage{hyperref}

\newcommand{\la}{\langle}
\newcommand{\ra}{\rangle}

\newcommand{\bbm}{\begin{bmatrix}}
\newcommand{\ebm}{\end{bmatrix}}
\newcommand{\bBm}{\begin{Bmatrix}}
\newcommand{\eBm}{\end{Bmatrix}}
\newcommand{\bpm}{\begin{pmatrix}}
\newcommand{\epm}{\end{pmatrix}}

\begin{document}


\title{Isospin splitting of the Dirac mass probed by the relativistic Brueckner-Hartree-Fock theory in the full Dirac space}

\author{Pianpian Qin}
\affiliation{Department of Physics, Chongqing Key Laboratory for Strongly Coupled Physics, Chongqing University, Chongqing 401331, China}

\author{Qiang Zhao}
\affiliation{Center for Exotic Nuclear Studies, Institute for Basic Science, \\
Daejeon 34126, Korea}

\author{Hui Tong}
\affiliation{Helmholtz-Institut f$\ddot{u}$r Strahlen- und Kernphysik and Bethe Center for Theoretical Physics, Universit$\ddot{a}$t Bonn, D-53115 Bonn, Germany}

\author{Chencan Wang}
\affiliation{School of Physics and astronomy, Sun Yat-Sen University, Zhuhai 519082, China}

\author{Sibo Wang}
\email{sbwang@cqu.edu.cn}
\affiliation{Department of Physics, Chongqing Key Laboratory for Strongly Coupled Physics, Chongqing University, Chongqing 401331, China}

\date{\today}

\begin{abstract}

The isospin splitting of the Dirac mass obtained with the relativistic Brueckner-Hartree-Fock (RBHF) theory is thoroughly investigated. 
From the perspective in the full Dirac space, the long-standing controversy between the momentum-independence approximation (MIA) method and the projection method on the isospin splitting of the Dirac mass in asymmetric nuclear matter (ANM) is analyzed in detail.
We find that, the \textit{assumption procedure} of the MIA method, which assumes that the single-particle potentials are momentum independent, is not a sufficient condition that directly leads to the wrong sign of the isospin splitting of the Dirac mass, while the \textit{extraction procedure} of the MIA method, which extracts the single-particle potentials from the single-particle potential energy, leads to the wrong sign. 
By approximately solving the set of equations involved in the \textit{extraction procedure}, a formal expression of the Dirac mass is obtained.
The wrong isospin splitting of the Dirac mass is mainly caused by that the \textit{extraction procedure} forcely assumes the momentum dependence of the single-particle potential energy to be a quadratic form where the strength is solely determined by the constant scalar potential.

\end{abstract}

\maketitle



It is still a challenging task to understand the properties of nuclear many-body system starting from the realistic nucleon-nucleon ($NN$) interaction.
The repulsive core of the realistic $NN$ interaction causes a strong correlation for the many-body wave function and requires advanced many-body methods that go beyond mean field~\cite{Baldo-2007-JPG.34.R243,Barrett2013_PPNP69-131,Hagen2014_RPP77-096302,Carlson2015_RMP87-1067,Hegert2016_PR621-165,SHEN-SH2019_PPNP109-103713}.
The Brueckner-Hartree-Fock (BHF) theory~\cite{Brueckner1954_PR95-217} is one of the representative nuclear many-body methods, which is characteristic for its capacity to soften the realistic $NN$ interaction to an effective $G$ matrix in nuclear medium.
The BHF theory can be derived as the two-hole-line truncation to the general Bethe-Brueckner-Goldstone expansion theory~\cite{Day1967RMP,Baldo-2007-JPG.34.R243}, where the ground-state properties of the nuclear many-body system are calculated order by order according to the number of independent hole lines contained in the expansion diagrams~\cite{Day1967RMP}.
Dating back to the 1960s, it is found that the saturation points of symmetric nuclear matter (SNM) calculated by the BHF theory with different two-body interactions are located on a so-called Coester line~\cite{Coester1970PRC}, which deviates systematically from the empirical values.
Within the nonrelativistic framework, a way out of this dilemma is to introduce three-body forces~\cite{1989-Grange-PhysRevC.40.1040, 1997-Baldo-AA328.274, ZuoW-2002-Nucl.Phys.A706.418, 2004-ZhouXR-PhysRevC.69.018801, Li-ZH_2006-PhysRevC.74.047304, LI-ZH2008_PRC77-034316}.


In 1980, major modification of the saturation properties of SNM is obtained by including a relativistic description of the nucleon motion~\cite{Anastasio1980_PRL45-2096}.
With this pioneering work, much attention has been paid to developing the relativistic Brueckner-Hartree-Fock (RBHF) theory~\cite{Anastasio1981_PRC23-2273, Brockmann1984-Phys.Lett.B, Horowitz1987_NPA464-613, Haar-1987-PRL59.1652}.
Starting from the Bonn potential~\cite{Machleidt1989_ANP19-189}, the saturation points of SNM obtained by the RBHF theory have been shifted remarkably close towards the empirical values, without introducing explicit three-body forces.
The success of the RBHF theory has been understood by the fact that, relativistic effects contribute a particular part of the three-body force~\cite{ZuoW-2002-Nucl.Phys.A706.418,Li-ZH_2006-PhysRevC.74.047304,Sammarruca_2012-PhysRevC.86.054317} through virtual nucleon-antinucleon excitations in the intermediate states (the so-called $Z$ diagrams)~\cite{Jaroszewicz1991_PRC43-1946, Brown-1987-Phys.Scr.}.


One of the essential points of the RBHF theory is using the Dirac equation to describe the single-nucleon motion in the mean field, i.e., the so-called single-particle potentials (SPPs).
The SPP operator is generally divided into a scalar and a vector component~\cite{Serot1986_ANP16-1}. They can be obtained in a self-consistent way from the effective $G$ matrix. 
In principle, the calculations of SPPs and the $G$ matrix should be performed in the full Dirac space, where both the positive-energy states (PESs) and negative-energy states (NESs) are included.
However, due to the complexity of these procedures, RBHF calculations have been usually performed without NESs, by invoking approximate methods to accomplish the Hartree-Fock calculation, such as the momentum-independence approximation (MIA) method~\cite{Brockmann1990_PRC42-1965,Sammarruca_2012-PhysRevC.86.054317,LI-GQ1992_PRC45-2782,Alonso-2003-PRC67.054301,TONG-H2018_PRC98-054302,WangCC-2020-JPG47.105108} and the projection method~\cite{Gross-Boelting1999_NPA648-105, Haar-1987-PRL59.1652,VanDalen-2004-NPA744.227,VanDalen-2005-Phys.Rev.Lett.95.022302,VanDalen-2005-PRC72.065803,VanDalen-2007-Eur.Phys.J.A31-29}.


The MIA method assumes that the SPPs are independent of the momentum, and extracts the SPPs from the single-particle potential energies calculated at two selected momenta.
In the projection method, the $G$ matrix elements are projected onto a complete set of Lorentz invariant amplitudes~\cite{Horowitz1987_NPA464-613}, from which the SPPs are calculated.
Since the $G$ matrix coupled to the NESs is not considered, the SPPs obtained with these two methods are both suffered from ambiguities~\cite{Nuppenau-1989-Nucl.Phys.A504839, Gross-Boelting1999_NPA648-105, WANG-SB2021_PRC103-054319}.
Moreover, two methods deviate significantly for the isovector properties in asymmetric nuclear matter (ANM), especially for the isospin splitting of the Dirac mass $M^*_{D,n} - M^*_{D,p}$~\cite{Ulrych1997_PRC56-1788,VanDalen-2005-Phys.Rev.Lett.95.022302,VanGiai2010_JPG37-064043}.
The MIA method leads to $M^*_{D,n} - M^*_{D,p} > 0$, while the projection method obtains the opposite sign $M^*_{D,n} - M^*_{D,p} < 0$.


Recently, self-consistent RBHF calculations in the full Dirac space have been achieved~\cite{WANG-SB2021_PRC103-054319,WANG-SB2022_PhysRevC.105.054309, 2022-Wang-SIBO-PhysRevC.106.L021305}, which avoid the ambiguities suffered from the RBHF calculations without NESs. In ANM, the full solution predicts the sign $M^*_{D,n}-M^*_{D,p}<0$~\cite{2022-Wang-SIBO-PhysRevC.106.L021305} and clarifies the long-standing controversy between the MIA method and the projection method on the isospin splitting of the Dirac mass.
The RBHF theory in the full Dirac space has also been applied to study the nonrelativistic effective mass in nuclear matter~\cite{SiboWang2023-PhysRevC.108.L031303}, the properties of $^{208}$Pb with a liquid droplet model~\cite{2022-TongH-PhysRevC.107.034302}, and the neutron star properties~\cite{Tong_2022-AstrophysicsJ930.137, 2022-Wang-PhysRevC.106.045804, 2023-QuXY-SciChina}.
In this work, we aim to thoroughly study the isospin splitting of the Dirac mass obtained with the RBHF calculations, especially for the performance of the MIA method, from the perspective in the full Dirac space.




In the RBHF theory, the single-particle motion for a nucleon inside the infinite nuclear matter is described by the Dirac equation
\begin{equation}\label{eq:DiracEquation}
  \left\{ \bm{\alpha}\cdot\bm{p}+\beta \left[M+\mathcal{U}_\tau(\bm{p})\right]\right\} u_\tau(\bm{p},s)
  = E_{\bm{p},\tau}u_{\tau}(\bm{p},s).
\end{equation}
Here $\bm{\alpha}$ and $\beta$ are the Dirac matrices, $M$ is the nucleon mass, $\bm{p}$ and $E_{\bm{p},\tau}$ are the momentum and single-particle energy, $s$ denotes the spin, and $\tau=n,p$ denotes neutron $n$ and proton $p$.
The symbol $u_\tau$ in Eq.~\eqref{eq:DiracEquation} represents the positive-energy Dirac spinor, while the negative-energy Dirac spinor $v_\tau$ is obtained by $v_\tau = \gamma_5 u_\tau$.
The SPP operator $\mathcal{U}_\tau(\bm{p})$ can be decomposed into the scalar potential $U_{S,\tau}(p)$, the timelike and spacelike parts of the vector potential $U_{0,\tau}(p)$ and $U_{V,\tau}(p)$~\cite{Serot1986_ANP16-1} as
\begin{equation}\label{eq:SPP}
  \mathcal{U}_\tau(\bm{p}) = U_{S,\tau}(p)+ \gamma^0U_{0,\tau}(p) + \bm{\gamma\cdot\hat{p}}U_{V,\tau}(p).
\end{equation}
Here $\hat{\bm{p}}=\bm{p}/|\bm{p}|=\bm{p}/p$ is the unit vector parallel to the momentum $\bm{p}$.
By calculating the matrix elements of $\mathcal{U}_\tau(\bm{p})$ on the basis expanded by the PESs and NESs, i.e., $\Sigma^{++}_\tau(p)$, $\Sigma^{-+}_\tau(p)$, and $\Sigma^{--}_\tau(p)$, the momentum dependent SPPs can be determined uniquely through~\cite{WANG-SB2021_PRC103-054319,Tong_2022-AstrophysicsJ930.137}
\begin{subequations}\label{eq:Sigma2US0V}
  \begin{align}
    U_{S,\tau}(p) = &\ \frac{\Sigma^{++}_\tau(p)-\Sigma^{--}_\tau(p)}{2}, \label{eq:Sigma2US0V-US}\\
    U_{0,\tau}(p) = &\ \frac{E^*_{\bm{p},\tau}}{M^*_{\bm{p},\tau}}\frac{\Sigma^{++}_\tau(p)+\Sigma^{--}_\tau(p)}{2}
               - \frac{p^*_\tau}{M^*_{\bm{p},\tau}}\Sigma^{-+}_\tau(p),  \label{eq:Sigma2US0V-U0}\\
    U_{V,\tau}(p) = &\ -\frac{p^*_\tau}{M^*_{\bm{p},\tau}}\frac{\Sigma^{++}_\tau(p)+\Sigma^{--}_\tau(p)}{2}
              + \frac{E^{*}_{\bm{p},\tau}}{M^*_{\bm{p},\tau}} \Sigma^{-+}_\tau(p).  \label{eq:Sigma2US0V-UV}
  \end{align}
\end{subequations}
Here the plus and minus signs in the superscripts denote the PESs and NESs, respectively.
The effective quantities in Eq.~\eqref{eq:Sigma2US0V} are defined as $\bm{p}^*_\tau = \bm{p}+\hat{\bm{p}}U_{V,\tau}(p)$,\ $M^*_{\bm{p},\tau} = M+U_{S,\tau}(p)$,\ and $E^*_{\bm{p},\tau} = E_{\bm{p},\tau}-U_{0,\tau}(p)$.
The effective mass $M^*_{\bm{p},\tau}$ is also known as the Dirac mass $M^*_{D,\tau}$.

The quantities $\Sigma^{++}_\tau(p)$, $\Sigma^{-+}_\tau(p)$, and $\Sigma^{--}_\tau(p)$ in Eq.~\eqref{eq:Sigma2US0V} are calculated by summing up the effective $G$ matrix elements with all the nucleons inside the Fermi sea
\begin{subequations}\label{eq:Gm2Sigma}
  \begin{align}
    \Sigma^{++}_\tau(p) = &\ \sum_{s'\tau'} \int^{k^{\tau'}_F}_0 \frac{d^3p'}{(2\pi)^3}
          \frac{M^*_{\bm{p}',\tau'}}{E^*_{\bm{p}',\tau'}}
          \la \bar{u}_\tau(\bm{p},1/2) \bar{u}_{\tau'}(\bm{p}',s')| \bar{G}^{++++}(W)|
          u_\tau(\bm{p},1/2)u_{\tau'}(\bm{p}',s')\ra, \label{eq:Sigma++} \\
    \Sigma^{-+}_\tau(p) = &\ \sum_{s'\tau'} \int^{k^{\tau'}_F}_0 \frac{d^3p'}{(2\pi)^3}
          \frac{M^*_{\bm{p}',\tau'}}{E^*_{\bm{p}',\tau'}}
          \la \bar{v}_\tau(\bm{p},1/2) \bar{u}_{\tau'}(\bm{p}',s')| \bar{G}^{-+++}(W)|
          u_\tau(\bm{p},1/2)u_{\tau'}(\bm{p}',s')\ra, \label{eq:Sigma-+} \\
    \Sigma^{--}_\tau(p) = &\ \sum_{s'\tau'} \int^{k^{\tau'}_F}_0 \frac{d^3p'}{(2\pi)^3}
          \frac{M^*_{\bm{p}',\tau'}}{E^*_{\bm{p}',\tau'}}
          \la \bar{v}_\tau(\bm{p},1/2) \bar{u}_{\tau'}(\bm{p}',s')| \bar{G}^{-+-+}(W)|
          v_\tau(\bm{p},1/2)u_{\tau'}(\bm{p}',s')\ra. \label{eq:Sigma--}
  \end{align}
\end{subequations}
Here $\bar{G}$ is the antisymmetrized $G$ matrix, $k^\tau_F$ is the Fermi momentum, and the starting energy is denoted by $W$.
The additional factors $M^*/E^*$ in Eq.~\eqref{eq:Gm2Sigma} arise due to the normalization of the Dirac spinors, i.e., $\bar{u}u=1, \bar{v}v=-1$.
The effective interaction $G$ matrix is the solution of the in-medium Thompson equation~\cite{Brockmann1990_PRC42-1965} which describes the two-body scattering
\begin{equation}\label{eq:ThomEqu}
  \begin{split}
  G_{\tau\tau'}(\bm{q}',\bm{q}|\bm{P},W)
  =&\ V_{\tau\tau'}(\bm{q}',\bm{q}|\bm{P})
  + \int \frac{d^3k}{(2\pi)^3}
  V_{\tau\tau'}(\bm{q}',\bm{k}|\bm{P}) \\
    & \times \frac{M^{*}_{\bm{P}+\bm{k},\tau}M^{*}_{\bm{P}-\bm{k},\tau'}}{E^*_{\bm{P}+\bm{k},\tau}E^*_{ \bm{P}-\bm{k},\tau'}}
    \frac{Q_{\tau\tau'}(\bm{k},\bm{P})}{W-E_{\bm{P}+\bm{k},\tau}-E_{\bm{P}-\bm{k},\tau'}}  G_{\tau\tau'}(\bm{k},\bm{q}|\bm{P},W).
  \end{split}
\end{equation}
The labels of PESs and NESs have been suppressed.
In this work, the realistic $NN$ interaction $V_{\tau\tau'}$ is chosen as the Bonn potential~\cite{Machleidt1989_ANP19-189}.
Here $\bm{P}=\frac{1}{2}({\bm k}_1+{\bm k}_2)$ and $\bm{k}=\frac{1}{2}({\bm k}_1-{\bm k}_2)$ are the center-of-mass and the relative momenta of the two interacting nucleons with momenta ${\bm k}_1$ and ${\bm k}_2$.
The initial, intermediate, and final relative momenta of the two nucleons are $\bm{q}, \bm{k}$, and $\bm{q}'$, respectively.
The $NN$ scattering in the nuclear medium is restricted with the Pauli operator $Q_{\tau\tau'}(\bm{k},\bm{P})$.

Equations~\eqref{eq:DiracEquation}, \eqref{eq:Sigma2US0V}, \eqref{eq:Gm2Sigma}, and \eqref{eq:ThomEqu} constitute a coupled system that has to be solved in a self-consistent way. After the convergence of SPPs, the single-particle and bulk properties of nuclear matter can be calculated straightforwardly~\cite{VanDalen-2004-NPA744.227,Katayama2013_PRC88-035805,TONG-H2018_PRC98-054302}.


\begin{figure}[htbp]
  \centering
  \includegraphics[width=10.0cm]{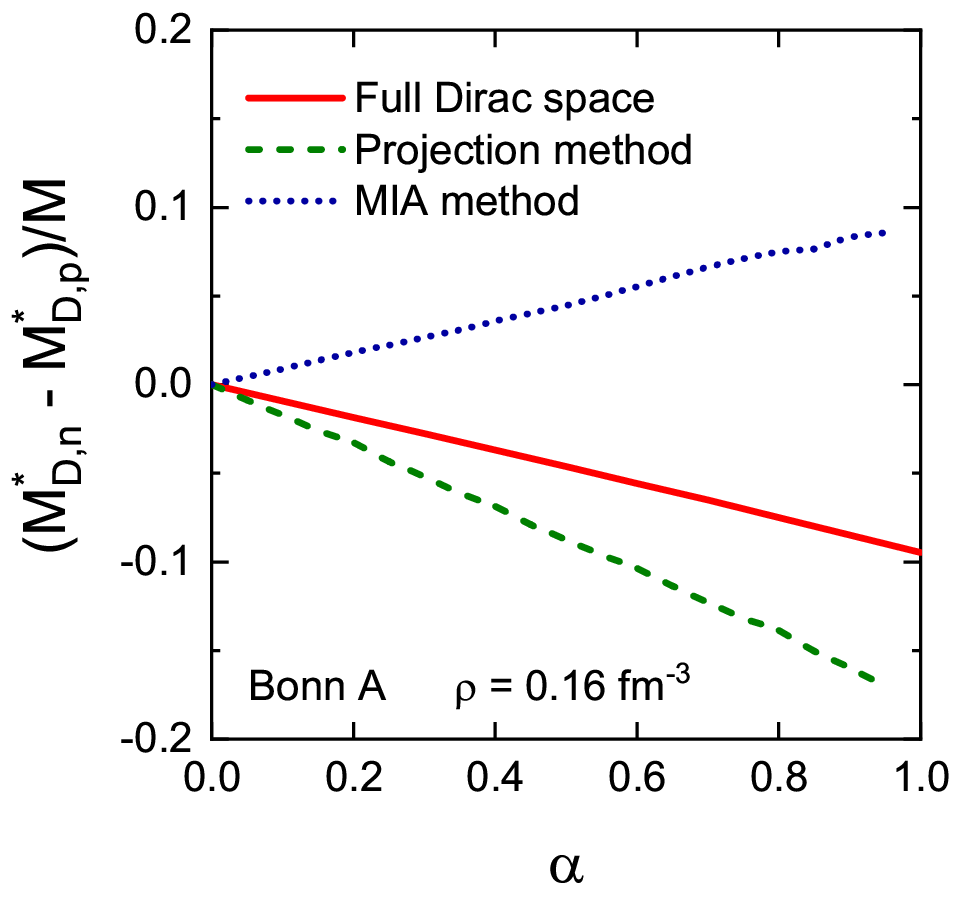}
  \caption{(Color online) The isospin splittings of Dirac masses $(M^*_{D,n}-M^*_{D,p})/M$ as functions of the asymmetry parameter $\alpha$ obtained by the RBHF theory in the full Dirac space (red solid), the projection method (olive dashed), and the MIA method (blue dotted).
  The density is fixed at the normal nuclear saturation density $\rho=0.16\ \text{fm}^{-3}$. The $NN$ interaction Bonn A is used.}
  \label{Fig1}
\end{figure}

In SNM, the Dirac masses for nucleon calculated by the MIA method and the projection method are both quantitatively close to the results obtained in the full Dirac space~\cite{2022-Wang-SIBO-PhysRevC.106.L021305}.
However, the situation changes dramatically in ANM.
Figure~\ref{Fig1} depicts the isospin splitting of the Dirac mass $(M^*_{D,n}-M^*_{D,p})/M$ calculated with the RBHF theory as a function of the asymmetry parameter $\alpha=(\rho_n-\rho_p)/\rho$.
The projection method finds that in ANM there is $M^*_{D,n}-M^*_{D,p}<0$, while the MIA method leads to the opposite sign $M^*_{D,n}-M^*_{D,p}>0$.
This contradiction, which is well known since 1997~\cite{Ulrych1997_PRC56-1788}, has been clarified recently by considering the PESs and the NESs simultaneously in Ref.~\cite{2022-Wang-SIBO-PhysRevC.106.L021305}.
Comparing to the RBHF results calculated in the full Dirac space as also shown in Fig.~\ref{Fig1}, the projection method obtains a qualitatively consistent isospin dependence of the Dirac mass with overestimated amplitudes, while the MIA method gives a wrong sign.

It is still unclear why the MIA method succeeds in SNM but fails in ANM.
To reach the answer, we notice that this method has two essential procedures. The first procedure can be called as the \textit{assumption procedure}, which assumes that the scalar potential and the timelike part of the vector potential are momentum independent and the spacelike part of the vector potential is negligible, i.e.,
\begin{equation}\label{eq-assume}
    U_{S,\tau}(p) \approx U_{S,\tau},\quad U_{0,\tau}(p) \approx U_{0,\tau}, \quad U_{V,\tau}(p) \approx 0.
\end{equation}
The second procedure can be called as the \textit{extraction procedure}, which extracts the two constants $U_{S,\tau}$ and $U_{0,\tau}$ from the single-particle potential energy $U_{\tau}(k)$ at two momenta $(k^\tau_1, k^\tau_2)$
\begin{subequations}\label{eq-extract}
  \begin{align}
      U_\tau(k^\tau_1) =&\ \frac{M+U_{S,\tau}}{\sqrt{(k^{\tau}_1)^2 + (M+U_{S,\tau})^2}} U_{S,\tau} 
          + U_{0,\tau},  \label{eq-extract-1}\\
      U_\tau(k^\tau_2) =&\ \frac{M+U_{S,\tau}}{\sqrt{(k^{\tau}_2)^2 + (M+U_{S,\tau})^2}} U_{S,\tau} 
          + U_{0,\tau}.  \label{eq-extract-2}
  \end{align} 
\end{subequations}
The quantity $U_{\tau}(k)$ in Eq.~\eqref{eq-extract} is calculated as $(M^*_{\tau}/E^*_{\bm{k},\tau})\Sigma^{++}_{\tau}(k)$.




\begin{figure}[htbp]
  \centering
  \includegraphics[width=10.0cm]{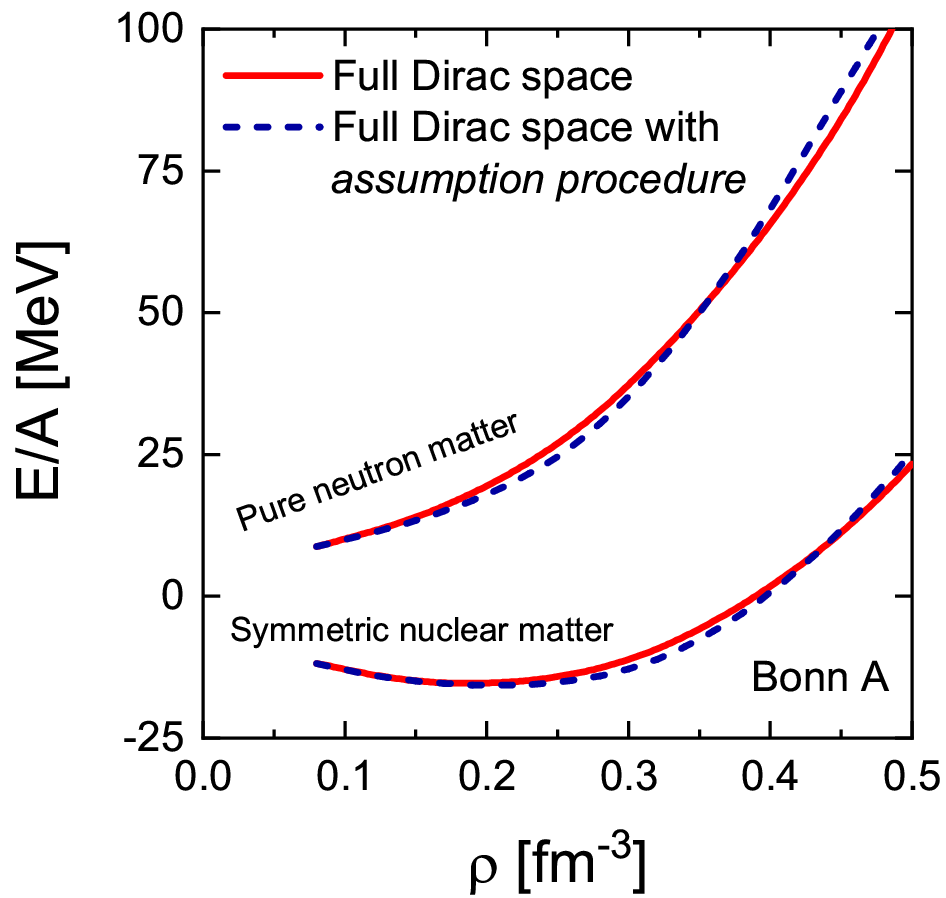}
  \caption{(Color online) Binding energies per nucleon $E/A$ of SNM and PNM as functions of the nucleon density $\rho$ calculated in the full Dirac space with the \textit{assumption procedure}. For comparison, the self-consistent results obtained by the RBHF theory in the full Dirac space are also shown.}
  \label{Fig2}
\end{figure}

The RBHF calculations in the full Dirac space provide us an opportunity to analyze in detail the isospin splitting of the Dirac mass.
In the following, we try to study further by testing separately the two procedures of the MIA method from the perspective in the full Dirac space. 

Firstly, we apply the \textit{assumption procedure} \eqref{eq-assume} in the full Dirac space, i.e., the quantities $U_{S,\tau}(p)$ in Eq.~\eqref{eq:Sigma2US0V-US} and $U_{0,\tau}(p)$ in Eq.~\eqref{eq:Sigma2US0V-U0} are assumed to be momentum independent and are calculated at the Fermi momentum $k^\tau_F$, while $U_{V,\tau}(p)$ in Eq.~\eqref{eq:Sigma2US0V-UV} is set as zero.
The newly obtained quantities $U_{S,\tau}$ and $U_{0,\tau}$ will be used to update the Dirac spinors and $G$ matrix for next iteration.
After the convergence of SPPs, we calculate the binding energies per nucleon for SNM and pure neutron matter (PNM). As can be seen in Fig.~\ref{Fig2}, the $E/A$ for SNM and PNM are very close to those obtained by the RBHF theory in the full Dirac space.
This indicates that the \textit{assumption procedure} is reasonable for describing the bulk properties of nuclear matter.


\begin{figure}[htbp]
  \centering
  \includegraphics[width=14.0cm]{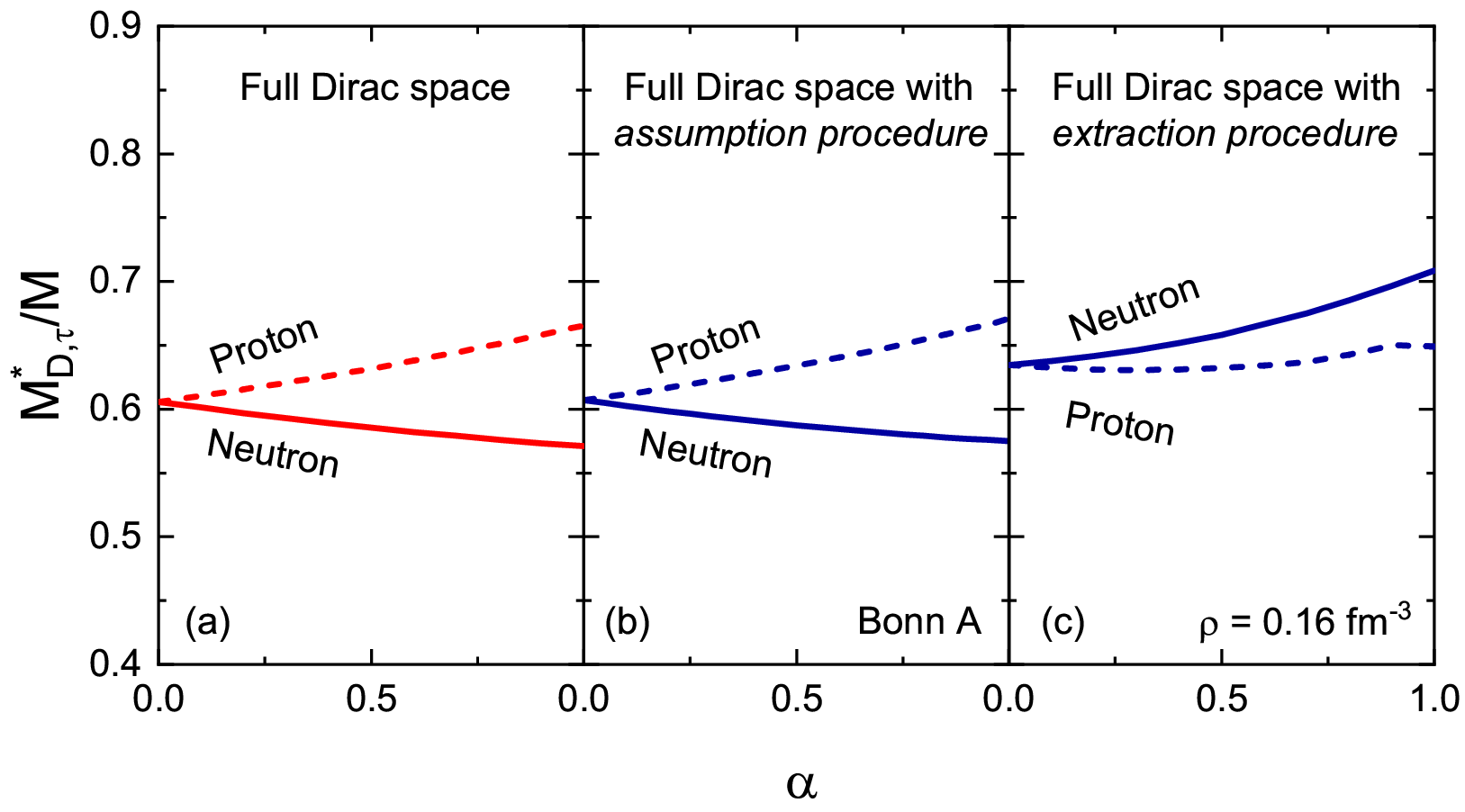}
  \caption{(Color online) Dirac masses for neutron (solid lines) and proton (dashed lines) as functions of the asymmetric parameter $\alpha$ at density $\rho=0.16\ \text{fm}^{-3}$ calculated by the RBHF theory (a) in the full Dirac space, (b) with the \textit{assumption procedure}, and (c) with the \textit{extraction procedure}. Details can be found in the text.}
  \label{Fig3}
\end{figure}

In Fig.~\ref{Fig3}(b), we show the Dirac masses of the neutron and proton obtained in the full Dirac space with the \textit{assumption procedure}.
The density is fixed at $\rho=0.16\ \text{fm}^{-3}$.
The relation of $M^*_{D,n}-M^*_{D,p}<0$ is found, which is consistent with the result obtained by the RBHF theory in the full Dirac space, as shown in Fig.~\ref{Fig3}(a).
This indicates that the \textit{assumption procedure} is not a sufficient condition that directly leads to the wrong sign when one calculates the isospin splitting of the Dirac mass in ANM.

Secondly, we test the influence of the \textit{extraction procedure} \eqref{eq-extract} from the perspective in the full Dirac space. Starting from the converged $U_{\tau}(p)$ obtained by the RBHF theory in the full Dirac space, we extract the two constants $U_{S,\tau}$ and $U_{0,\tau}$ by using Eq.~\eqref{eq-extract} with two selected momenta $(0.7k^\tau_F,k^\tau_F)$ and then calculate the Dirac mass. 
The resulting isospin splitting of the Dirac mass is shown in Fig.~\ref{Fig3}(c).
It is seen that $M^*_{D,n}-M^*_{D,p}>0$ is obtained for the entire region of the asymmetry parameter $\alpha$, which is opposite to results obtained in the full Dirac space (Fig.~\ref{Fig3}(a)).
This shows that it is the \textit{extraction procedure} \eqref{eq-extract} that leads to the wrong sign of the isospin splitting of the Dirac mass.


\begin{figure}[htbp]
  \centering
  \includegraphics[width=16.0cm]{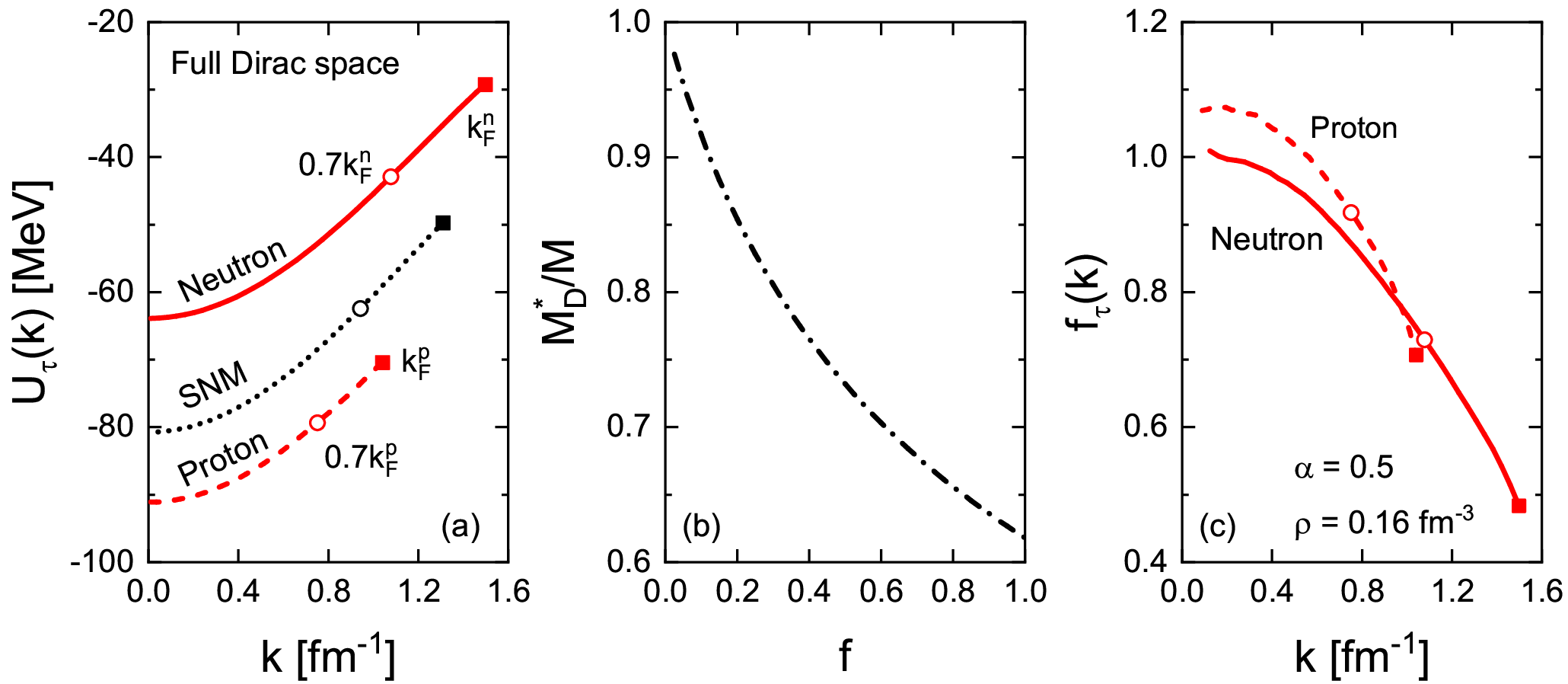}
  \caption{(Color online)
  (a) The single-particle potential energy $U_\tau(k)$ obtained in the full Dirac space as a function of momentum $k$. The cases for SNM and ANM with $\alpha=0.5$ and $\rho=0.16\ \text{fm}^{-3}$ are shown.
  The momenta $0.7k^\tau_F$ and $k^\tau_F$ are highlighted with empty dots and solid squares respectively.
  (b) The Dirac mass $M^*_D/M$ as a functional of dimensionless quantity $f$ in Eq.~\eqref{eq-sol-snm}.
  (c) The dimensionless function $f$ \eqref{eq:fdef} obtained in the full Dirac space as a function of momentum $k$. The notation for lines and symbols is the same as panel (a).}
  \label{Fig4}
\end{figure}

To further investigate how the wrong isospin dependence of the Dirac mass emerges from the \textit{extraction procedure}, it is insightful to solve Eq.~\eqref{eq-extract} to get a formal expression of the Dirac mass. 
We start from the case of SNM and suppress the isospin indexes for moment.
For $M\simeq 1000$ MeV, $U_S\simeq -400$ MeV, and $k_F = 1.34\ \text{fm}^{-1}$, one notices that the typical choice $k_1 = 0.7k_F$ and $k_2=k_F$ leads to $\left[k_1/(M+U_S)\right]^2\simeq 0.1\ll 1$, $\left[k_2/(M+U_S)\right]^2\simeq 0.2\ll1$.
This allows one to expand the square root in Eq.~\eqref{eq-extract} to the first order,
\begin{equation}\label{eq:quad}
  U(k) = -\frac{1}{2}\frac{U_S}{(M+U_S)^2} k^2 + U_0,
\end{equation}
which indicates a quadratic form of the single-particle potential energy $U(k)$.
With Eq.~\eqref{eq:quad} it is straightforward to find the difference between $U(k_2)$ and $U(k_1)$ as
\begin{equation}\label{eqU2mU1-sim}
    U(k_2) - U(k_1) = -\frac{1}{2}\frac{U_S}{(M+U_S)^2} \left( k^2_2 - k^2_1\right).
\end{equation}
For simplicity, we consider the limit that two momenta $k_1$ and $k_2$ are very close and denote them as $k$.
In this case Eq.~\eqref{eqU2mU1-sim} can be written as a quadratic equation for $M^*_D/M$
\begin{equation}\label{eqMUs}
   f(k) (M^*_D/M)^2 + M^*_D/M - 1 = 0,
\end{equation}
where the dimensionless function $f(k)$ is defined as
\begin{equation}\label{eq:fdef}
  f(k) \equiv \lim_{ \substack{k_2\rightarrow k \\ k_1\rightarrow k}} 2M\frac{U(k_2) - U(k_1)}{k^2_2-k^2_1} 
    = M\frac{U'(k)}{k}.
\end{equation}
The derivative $U'(k)$ describes the momentum dependence of the single-particle potential energy $U(k)$.

In Fig.~\ref{Fig4}(a), the single-particle potential energy $U_\tau(k)$ obtained in the full Dirac space for SNM and also ANM with $\alpha=0.5$ and $\rho=0.16\ \text{fm}^{-3}$ is shown as a function of momentum $k$. It is found that $U_\tau(k)$ is a monotonic function of the $k$. Therefore, the function $f(k)$ is positive definite and the solution of Eq.~\eqref{eqMUs} is  
\begin{equation}\label{eq-sol-snm}
  M^*_D/M = \frac{\sqrt{1+4 f(k)}-1}{2 f(k)}.
\end{equation}
From Fig.~\ref{Fig4}(b), the Dirac mass $M^*_D/M$ is decreasing with increasing $f$.
Therefore, the sign of $M^*_{D,n} - M^*_{D,p}$ is opposite to that of $f_n(k_n) - f_p(k_p)$, i.e.,
\begin{equation} \label{eq-MnMpif}
  \begin{cases}
    M^*_{D,n} - M^*_{D,p} < 0, \quad \text{if}\quad f_n(k_n) > f_p(k_p),\\
    M^*_{D,n} - M^*_{D,p} > 0, \quad \text{if}\quad f_n(k_n) < f_p(k_p).
  \end{cases}
\end{equation}
In Fig.~\ref{Fig4}(c), the function $f_\tau(k)$ for nucleon $\tau$ obtained in the full Dirac space is shown.
It is found that $f_\tau(k)$ is decreasing with the increasing $k$.
Besides, generally there is $f_n(k)<f_p(k)$ for $k<k^p_F$. This indicates that the momentum dependence of $U_n(k)$ is weaker than that of $U_p(k)$. 

In practice, one usually chooses $k^\tau_1 = 0.7k^\tau_F$ and $k^\tau_2 = k^\tau_F$~\cite{TONG-H2018_PRC98-054302,2022-Wang-SIBO-PhysRevC.106.L021305}. For ANM with $\alpha>0$, the Fermi momentum for neutron $k^n_F$ is larger than the of proton, i.e., $k^n_F>k^p_F$. As shown by the symbols in panels (a) and (c) in Fig.~\ref{Fig4}, this choice leads to $k_n>k_p$ and $f_n(k_n) < f_p(k_p)$.
Specifically, starting from the self-consistent $U_\tau(k)$ obtained in the full Dirac space, the choices $(0.7k^n_F, k^n_F)$ for neutron and $(0.7k^p_F, k^p_F)$ for proton lead to $f_n(k_n)\approx 0.6$ and $f_p(k_p)\approx 0.8$.
According to the analysis in Eq.~\eqref{eq-MnMpif}, the isospin splitting of the Dirac mass $M^*_{D,n} - M^*_{D,p}>0$ emerges.
This explains the reason why the MIA method leads to the wrong isospin splitting of the Dirac mass in ANM.

Come back to Eq.~\eqref{eq:quad}, one can see that the \textit{extraction procedure} forcely assumes that the momentum dependence of the single-particle potential energy to be a quadratic form, where the strength $-\frac{1}{2}\frac{U_S}{(M+U_S)^2}$ is solely determined by the scalar potential $U_S$.
In ANM, the momentum dependence of $U_n(k)$ is generally weaker than that of $U_p(k)$, as shown in Fig.~\ref{Fig4}(c).
It is exactly this fact that leads to $U_{S,n}>U_{S,p}$ and the wrong sign of the isospin splitting of the Dirac mass $M^*_{D,n}-M^*_{D,p}>0$.

In summary, the relativistic Brueckner-Hartree-Fock (RBHF) theory plays an important role in deriving the nuclear many-body properties from the realistic nucleon-nucleon interactions. In comparison to the results obtained self-consistently in the full Dirac space, the momentum-independence approximation (MIA) method leads to wrong isospin splitting of the Dirac mass in asymmetric nuclear matter (ANM).
The performance of this method is explored in detail from the perspective in the full Dirac space.
The \textit{assumption procedure} of the MIA method, which assumes that the single-particle potentials are momentum independent, is not a sufficient condition that directly leads to the wrong sign of the isospin splitting of the Dirac mass, while the \textit{extraction procedure} of the MIA method, which extracts the single-particle potentials from the single-particle potential energy, is found to be responsible for the wrong isospin splitting of the Dirac mass.
By approximately solving the set of equations involved in the \textit{extraction procedure}, a formal expression of the Dirac mass is obtained.
With the typical choice of momenta adopted in practical MIA calculations, the wrong isospin splitting of the Dirac mass is found.
We conclude that the wrong isospin splitting of the Dirac mass emerges from the fact that the \textit{extraction procedure} forcely assumes the momentum dependence of the single-particle potential energy to be a quadratic form where the strength is solely determined by the constant scalar potential. 
This work improves greatly our understanding on the isospin splitting of the Dirac mass with the RBHF theory.

\begin{acknowledgments}

The authors thank the discussion with Professor Peter Ring.
This work was supported in part by the China Postdoctoral Science Foundation under Grant No. 2021M700610, the National Natural Science Foundation of China (NSFC) under Grants No. 12205030, the Fundamental Research Funds for the Central Universities under Grants No. 2020CDJQY-Z003 and No. 2021CDJZYJH-003, and the Institute for Basic Science under Grant No. IBS-R031-D1. 
Part of this work was achieved by using the supercomputer OCTOPUS at the Cybermedia Center, Osaka University under the support of Research Center for Nuclear Physics of Osaka University and the High Performance Computing Resources in the Research Solution Center, Institute for Basic Science.

\end{acknowledgments}

\bibliography{MIA-FDS}

\end{document}